\documentclass[iop]{emulateapj}
\usepackage{amsmath,amssymb,times,graphics,color}

\newcommand{\ldl}{\ensuremath{\lambda/{\Delta}{\lambda}}}

\newcommand{\Teff}{T\ensuremath{_{\rm eff}}}

\newcommand{\ha}{H\ensuremath{\alpha}}
\newcommand{\Ki}{\ion{K}{1}}

\newcommand{\Nai}{\ion{Na}{1}}

\newcommand{\Lii}{\ion{Li}{1}}

\newcommand{\meth}{CH\ensuremath{_{4}}}
\newcommand{\wat}{H\ensuremath{_{2}}O}

\newcommand{\Hh}{H\ensuremath{_{2}}}
\newcommand{\kms}{km~s\ensuremath{^{-1}}}

\newcommand{\vtan}{V\ensuremath{_{\rm tan}}}

\newcommand{\chisq}{\ensuremath{\chi^2}}
\newcommand{\masyr}{mas~yr\ensuremath{^{-1}}}

\newcommand{\Msun}{M\ensuremath{_{\sun}}}
\newcommand{\Mtot}{M\ensuremath{_{\rm tot}}}
\newcommand{\JK}{\ensuremath{J-K_s}}

\newcommand{\name}{2MASS~J01303563$-$4445411}
\newcommand{\namesh}{2MASS~J0130$-$4445}

\shorttitle{M/L Dwarf Binary 2MASS~J01303563$-$4445411AB}
\shortauthors{Dhital et al.}

\begin{document}

\journalinfo{Accepted by the Astronomical Journal}
\accepted{October 6, 2010}
\submitted{Submitted: August 9, 2010}
\title{Resolved Spectroscopy of M Dwarf/L Dwarf
  Binaries. \\ IV. Discovery of an M9 + L6 Binary Separated by Over 100 AU} 
\author{
  Saurav Dhital       \altaffilmark{1},
  Adam J.\ Burgasser  \altaffilmark{2,5},
  Dagny L.\ Looper    \altaffilmark{3,5},
  Keivan G.\ Stassun  \altaffilmark{1,4}}
\altaffiltext{1}{Department of Physics \& Astronomy, Vanderbilt University,
  Nashville, TN, 37235, USA; saurav.dhital@vanderbilt.edu}
\altaffiltext{2}{Center of Astrophysics and Space Sciences, Department
  of Physics, University of California, San Diego, CA 92093, USA} 
\altaffiltext{3}{Institute for Astronomy, University of Hawaii, 2680
  Woodlawn Drive, Honolulu, HI 96822, USA} 
\altaffiltext{4}{Department of Physics, Fisk University, 1000 17th
  Avenue North, Nashville, TN 37208, USA}
\altaffiltext{5}{Visiting Astronomer at the Infrared Telescope
  Facility, which is operated by the University of Hawaii under
  Cooperative Agreement no. NCC 5-538 with the National Aeronautics
  and Space Administration, Science Mission Directorate, Planetary
  Astronomy Program.}

\begin{abstract}
We report the discovery of a faint L6$\pm$1 companion to the
previously known M9 dwarf, {\name}, based on our near-infrared imaging
and spectroscopic observations with the 3m Infrared Telescope Facility
SpeX imager/spectrometer. The visual binary is separated by
3$\farcs$28$\pm$0$\farcs$05 on the sky at a spectrophotometric
distance of 40$\pm$14~pc. The projected physical separation is
130$\pm$50~AU, making it one of the widest VLM field multiples
containing a brown dwarf companion. {\namesh} is only one of ten wide
VLM pairs and only one of six in the field. The secondary is considerably
fainter ($\Delta K \approx$ 2.35~mag) and redder ($\Delta$\ ({\JK})
$\approx$ 0.81~dex), consistent with component near-infrared types of
M9.0$\pm$0.5 and L6$\pm$1 based on our resolved spectroscopy. The
component types suggest a secondary mass below the
hydrogen-burning limit and an age-dependent mass ratio of
0.6--0.9. The system's space motion and spectroscopic indicators
suggest an age of 2--4~Gyr while the model-dependent masses and
binding energies suggest that this system is unlikely to have formed
via dynamical ejection. The age, composition, and separation of the
{\name} system make it useful for tests of VLM formation theories and
of condensate cloud formation in L dwarfs.
\end{abstract}


\section{Introduction}\label{Sec: intro}
The processes by which very low-mass (VLM; M $\lesssim$0.1
{\Msun}, \citealt{Burgasser2007b}) stars  and brown dwarfs (BD) form,
and whether these processes are similar to those of higher-mass stars,
is an open question. The VLMs/BDs exhibit significant differences in
the distribution of binary/multiple systems when compared to their
more massive brethren.  The resolved binary fraction of $\sim$20--30\%
in VLMs/BDs \citep{Basri2006, Joergens2008} is significantly lower
than in F and G dwarfs \citep[$\sim$60\%;][]{Duquennoy1991} and
modestly lower than M dwarfs \citep[$\sim$27--42\%;][]{Fischer1992,
  Reid1997}. The typical orbital separation  of $\sim$4--5~AU in VLMs/BDs
is much smaller compared to $\sim$30~AU for F, G, and M dwarf binaries
\citep{Duquennoy1991, Fischer1992}. In addition, while stellar
binaries are known to have separations in excess of $\sim$1~pc
\citep[e.g.,][]{Lepine2007a, Dhital2010}, no VLM system has a
separation greater than 6700~AU. Indeed, only 15 of the known 99 VLM
systems have projected physical separations larger than 20~AU and only
nine systems are wider than 100~AU\footnote{VLM Binaries Archive
  (\url{http://vlmbinaries.org/}) and references  therein.}. Energetically,
the VLM binaries seem to stand apart as well: based on empirical data,
\citet{Close2003} suggested minimum binding energies of
$10^{42.5}$~erg for field VLM systems, $\sim$300 times higher than the
$10^{40}$~erg limit for stellar binary systems. Lastly, most VLM
binaries are close to equal-mass. All of these differences indicate
that the same formation process(es) may not be responsible for the two
populations.

It is now generally believed that most stars form in multiple systems
via fragmentation of the protostellar cloud, with single stars being
the result of decay of unstable multiples \citep[e.g.,][]{Kroupa1995a}.  
The most favored process is gravoturbulence where the fragmentation is
the result of a combination of turbulent gas flows and gravity.
Hydrodynamical simulations have shown that when turbulent gas flows in
protostellar clouds collide, they form clumps that are gravitationally
unstable and, hence, collapse forming multiple stellar embryos 
\citep[e.g.,][]{Caselli2002, Goodwin2004a, Goodwin2004b, Bate2009}. 
Within a few freefall times, most of these embryos are ejected due to
mutual dynamical interactions, preferentially the ones with lower masses. 

To then explain the observed distributions of VLM binaries
separations, two explanations have been proffered. The first so-called
``ejection hypothesis'' suggests that most VLM binaries, unlike the
more-massive stellar systems, are the result of the ejected embryos
\citep{Reipurth2001}. The wider systems get disrupted, explaining the
overall rarity of VLM and BD binaries. The second is preferential
accretion within the first 0.1~Myr ($\sim$1 freefall time), making VLM
systems tighter and more equal-mass. As a result, even VLM
distributions that initially may have looked similar to that of higher
mass stars are transformed and look like the observed VLM
distributions \citep{Bate2009}. However, neither hypothesis explains
why $\sim$10\% of observed VLM binaries are wider than 100~AU. Two
other theories on VLM/BD formation, disk fragmentation
\citep[e.g.,][]{Watkins1998a, Watkins1998b} and photoablation
\citep{Whitworth2004}, require massive stars to trigger the process
and cannot explain the existence of VLM binaries in the
field. To resolve the differences between observational and numerical
results and to distinguish between the various formation scenarios, a
larger sample of VLM binaries---especially very wide systems that are
most susceptible to dynamical effects---is needed.

In this paper, we report the discovery of a wide VLM binary {\name}
(hereafter {\namesh}) separated by 130~AU, 3$\farcs$3. The brighter primary
component of {\namesh} was identified by \citet{Reid2008} in the Two
Micron All Sky Survey \citep[2MASS;][]{Skrutskie2006} and classified
as an M9 dwarf on the \citet{Kirkpatrick1999} red optical scheme,
indicating a spectrophotometric distance of 33.1$\pm$2.2~pc. Neither
{\ha} nor {\Lii}, activity and age indicators, respectively, were
evident in the optical spectrum.  The primary has a proper motion of
(120$\pm$14, -25$\pm$20)~\masyr and a tangential velocity of
19$\pm$3~\kms \citep{Faherty2009}. The system is unresolved in 2MASS,
and there have been no reports of a faint companion to this source in
either optical survey data or follow-up observations \citep{Reid2008,
  Faherty2009}.

In our own follow-up observations of {\namesh}, we have identified a
well-separated, faint L dwarf companion, indicating that this is a
wide VLM binary system with a probable BD component. In
Sections~\ref{Sec: NIRimaging} and \ref{Sec: NIRspectra}, we
describe our imaging and spectroscopic observations, respectively, and
discuss the properties of the components of the resolved binary system.
We discuss the physical association, mass, and age of the binary
{\namesh}AB in Section~\ref{Sec: analysis} and its implications on VLM
formation and evolution scenarios in Section~\ref{Sec: discussion}. 
The conclusions are presented in Section~\ref{Sec: summary}.
 
\section{Near-Infrared Imaging}\label{Sec: NIRimaging}
\subsection{Observations and Data Reduction}
{\namesh} was imaged with the 3m NASA Infrared Telescope Facility (IRTF)
SpeX spectrograph \citep{Rayner2003} on December 7, 2009 (UT), as part
of a program to identify unresolved M/L dwarf plus T dwarf spectral
binaries \citep[e.g.,][]{Burgasser2008a}.  Conditions were clear but
with poor seeing, 1$\farcs$2 at $K$-band, due in part to the large
airmass of the observation (2.34--2.37). These images revealed a faint
point source due east of the primary target at a separation of roughly
3$\arcsec$.  Four dithered exposures were obtained of the pair 
in each of the MKO\footnote{Mauna Kea Observatory filter system; see
  \citet{Tokunaga2002} and \citet{Simons2002}.} $J$, $H$, and $K$
filters, with individual exposure times of 45s, 30s, and 30s,
respectively. The field rotator was aligned at a position angle of
0$\degr$; i.e., north up and east to the left. 

\begin{figure}
  \begin{centering}
    \includegraphics[width=1\linewidth]{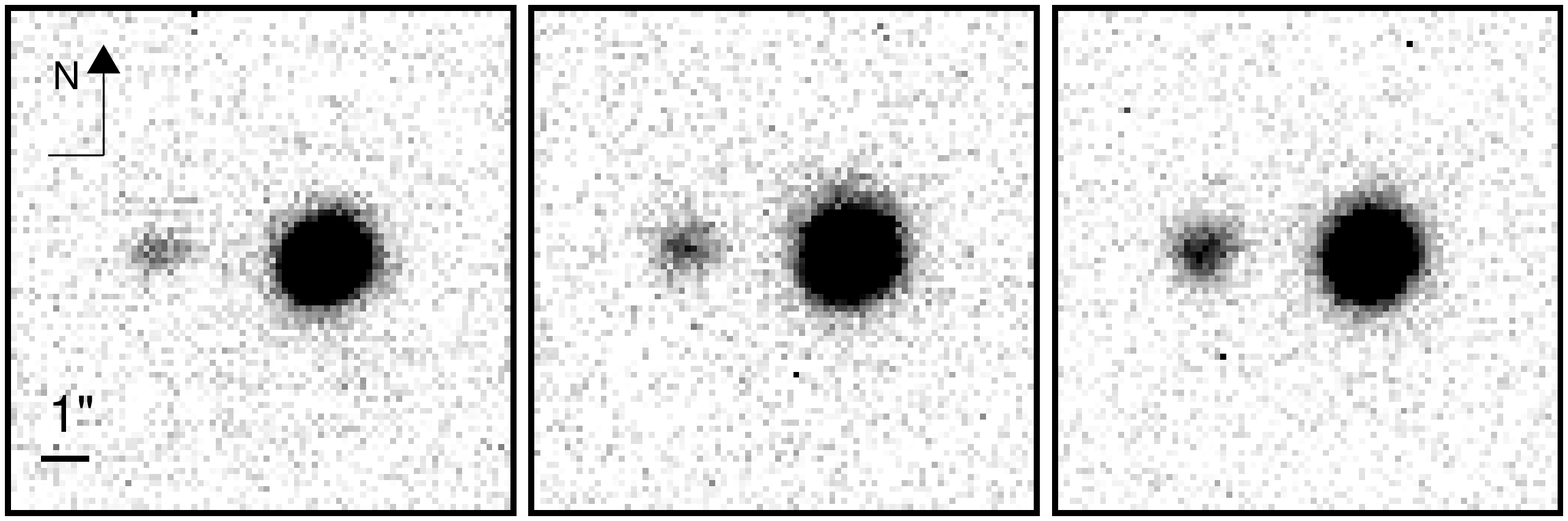}
    \caption{Combined SpeX images of {\namesh}AB 
      in the $J$-, $H$- and $K$-bands (left to right), showing the 
      10$\arcsec\times$10$\arcsec$ (83$\times$83 pixel) region
      around both sources.  Images are aligned with north up and
      east to the left.}
    \label{Fig: image}
  \end{centering}
\end{figure}

Imaging data were reduced in a standard manner using custom IDL routines.
Raw images were mirror-flipped about the y-axis to reproduce the sky
orientation and pair-wise subtracted to remove sky contributions. The
difference images were divided by normalized flat field frames,
constructed by median-combining the imaging data for each filter after
masking out the sources. Subsections of each image, 10$\arcsec$ (83
pixels) on a side and centered on the target source, were extracted
from these calibrated frames.  A final image for each filter/target
pair (Figure~\ref{Fig: image}) was produced by averaging the registered
subframes together, rejecting 5$\sigma$ pixel outliers.  The two
sources of {\namesh} are well resolved along a nearly east-west
axis. The brighter western component is hereafter referred to as
{\namesh}A and the eastern component as {\namesh}B.

\subsection{Analysis}
Component magnitudes and the angular separation of the {\namesh} pair
were determined through point spread function (PSF) fits to the
reduced imaging data, following the prescription described in
\citet{McElwain2006}.  The PSF models were derived from
Gaussian fits to the primary component in the individual subimage
frames.  For each filter, four distinct PSF models were produced, each
of which were fit to the individual images, resulting in a total of 16
independent measures of the relative component magnitudes and 48
independent measures of the separation and orientation of the
pair, in each of the $JHK$ filters. However, as the secondary was
undetected in one of the four $J$-band images, four measures of the
relative $J$-band flux and separation were discarded before computing
mean values and standard deviations. Separation measurements were
converted from pixels to arcseconds assuming a plate scale of 0$\farcs$120$\pm$0$\farcs$002
pixel$^{-1}$ (J.~Rayner, 2005, private communication) and no
distortion. The position angle (set at 0$\degr$) was assumed to be
accurate to within 0$\fdg$25 (ibid.).

Results are listed in Table~\ref{Tab: psf}. The angular separation of
the pair was measured to be 3$\farcs$282$\pm$0$\farcs$047 at a position
angle of 87$\fdg$3$\pm$0$\fdg$9; i.e., along an
east-west line. The secondary is both considerably fainter and significantly
redder than the primary. We derived relative magnitudes of $\Delta{J}=$
3.11$\pm$0.06 and $\Delta{K}=$ 2.34$\pm$0.04. Using the combined-light 2MASS 
photometry for the system\footnote{We included
  small corrections to the relative magnitudes in converting from the
  MKO to 2MASS photometric systems: 0.009, -0.006, and -0.003 mag in
  the $J$, $H$, and $K/K_s$ bands, respectively, calculated directly
  from the spectral data.}, this translates into {\JK} colors of
1.13$\pm$0.04 and 1.94$\pm$0.08 for the primary and secondary,
respectively.

\begin{deluxetable}{lc}
  \tablecaption{Results of PSF Fitting}
  \tablewidth{0pt}
  \tablehead{\colhead{Parameter} & \colhead{Value}}
  \startdata
  $\Delta{\alpha}\cos{\delta}$ ($\arcsec$)\tablenotemark{a} & 3.28$\pm$0.05 \\
  $\Delta{\delta}$ ($\arcsec$)\tablenotemark{a} & 0.15$\pm$0.06 \\
  Apparent Separation ($\arcsec$) & 3.28$\pm$0.05   \\
  Position Angle ($\degr$)\tablenotemark{a} & 87.3$\pm$0.9   \\
  $\Delta{J}$ (mag) & 3.11$\pm$0.06 \\
  $\Delta{H}$ (mag) & 2.68$\pm$0.11 \\
  $\Delta{K}$ (mag) & 2.34$\pm$0.04 \\
  \enddata
  \label{Tab: psf}
  \tablenotetext{a}{Measured from the brighter primary to the fainter secondary.}
\end{deluxetable}

\section{Near-Infrared Spectroscopy}\label{Sec: NIRspectra}
\subsection{Observations and Data Reduction}
The two components of {\namesh} were observed on separate nights with the  
prism-dispersed mode of SpeX, the primary on December 7, 2009 (the same
night as the imaging observations) and the secondary on December
28, 2009 (UT). Conditions on the latter night were clear with a stable
seeing of 0$\farcs$6 at $K$-band.
The SpeX prism mode provides 0.75--2.5~$\micron$ continuous
spectroscopy with resolution {\ldl} $\approx 120$ for the 0$\farcs$5
slit employed (dispersion across the chip is 20--30~{\AA}~pixel$^{-1}$).
Both components were observed separately, with the slit oriented
north-south, roughly aligned with the parallactic angle and 
perpendicular to the separation axis. 
 For the primary, eight exposures of 90s each were obtained at
an average airmass of 2.33, while guiding on spillover light from the
slit. For the secondary, eight exposures of 150s each were obtained at
an average airmass of 2.34, while guiding on the primary off-slit. For
both sources, the A0 V star HD~8977 was observed immediately before
the target for telluric and flux calibration while the quartz and Ar
arc lamps were observed for flat field and wavelength calibration,
respectively. Data were reduced using the SpeXtool package, version
3.4 \citep{Vacca2003, Cushing2004} using standard settings; see
\citet{Burgasser2007d} for details.

\subsection{Analysis}\label{Sec: NIRanalysis}
\begin{figure}
  \begin{centering}
    \includegraphics[width=0.8\linewidth]{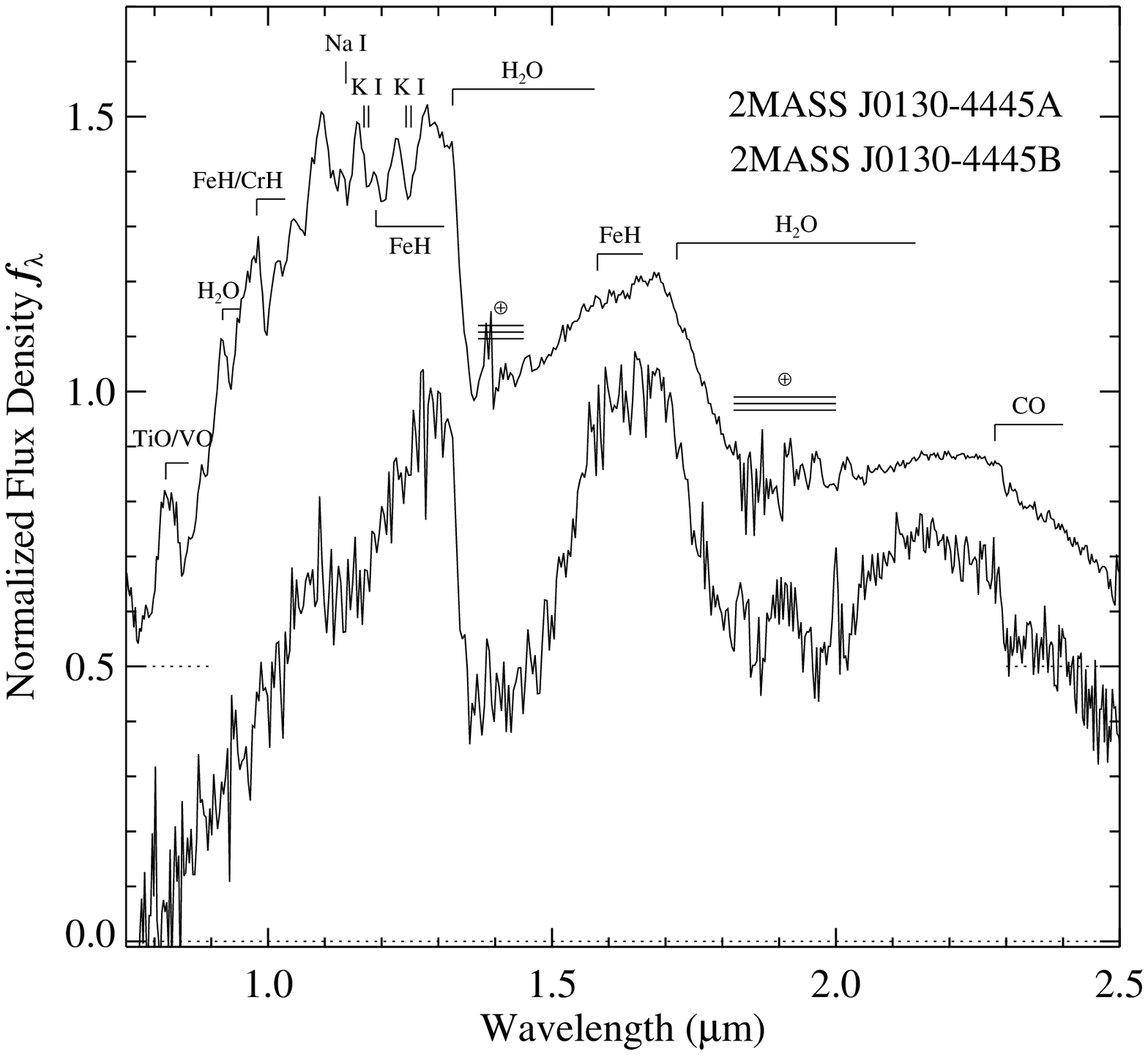}
    \caption{NIR spectra of {\namesh}A (top) and {\namesh}B
      (bottom) obtained with IRTF/SpeX. Data are normalized
      at the peak of each spectra, and the spectrum for {\namesh}A is
      vertically offset by 0.5~dex for clarity (dotted lines). NIR
      spectral features are labeled.}
    \label{Fig: nirspec}
  \end{centering}
\end{figure}

Figure \ref{Fig: nirspec} shows the spectra of the two components of
{\namesh}AB; signal-to-noise at the $JHK$ flux peaks was 100--150 and
25--35 for the A and B components, respectively. Both spectra show the
characteristic near-infrared (NIR) features of 
late-type M and L dwarfs \citep[e.g.,][]{Reid2001a, McLean2003,
  Cushing2005}: steep red optical slopes (0.8--1.0~$\micron$) from the
pressure broadened wing of the 0.77~$\micron$ {\Ki} doublet; molecular
absorption bands arising from {\wat} (1.4 and 1.9~$\micron$), CO
(2.3~$\micron$), and FeH (0.99~$\micron$); and an overall red spectral
energy distribution (SED), consistent with the photometric colors.
{\namesh}A also exhibits additional absorption features in the
0.8--1.2~$\micron$ region arising from TiO, VO, FeH, and unresolved {\Ki} and {\Nai}
lines, all typical for a late-type M dwarf. The corresponding region
in the spectrum of {\namesh}B is considerably smoother, albeit more
noisy, suggesting that many of these gaseous species have condensed out
\citep[e.g.,][]{Tsuji1998, Ackerman2001}. The appearance of weak
{\wat} absorption at 1.15~$\micron$ and the very red SED of the
NIR spectrum all indicate that {\namesh}B is a mid- to late-type L
dwarf with relatively thick condensate clouds at the photosphere. 

\begin{figure}
  \begin{centering}
    \includegraphics[width=1\linewidth]{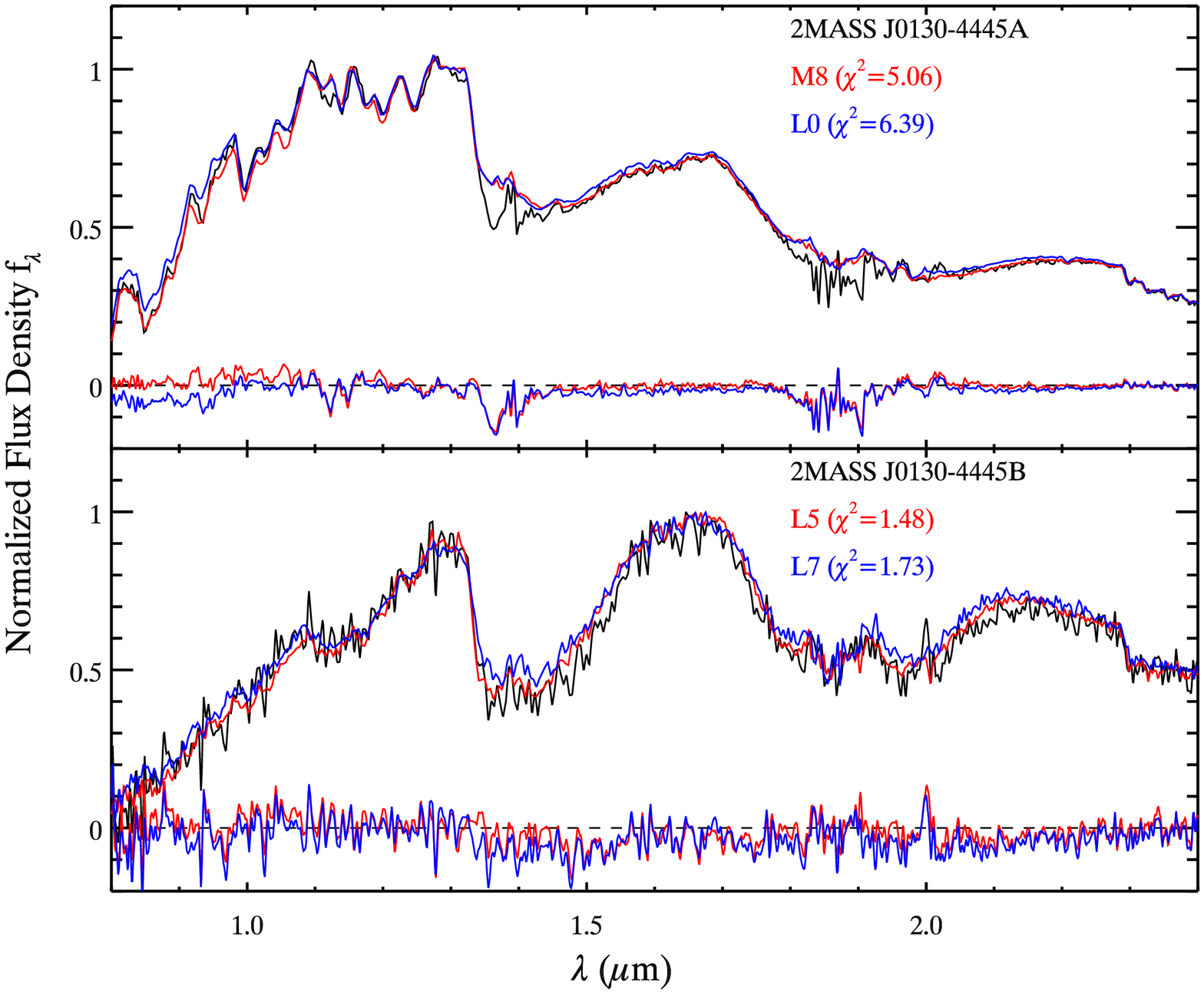}
    \caption{The spectral types for {\namesh}AB as determined by
      matching their spectra with templates from the SpeX Prism
      Spectral Libraries. The best matches for {\namesh}A were
      2MASS~J05173729$-$3348593 \citep[optically classified
      M8;][]{Cruz2003, Schmidt2007} and DENIS$-$P~J0652197$-$253450
      \citep[optically classified L0;][]{Phan-Bao2008} while
      2MASSW~J1326201$-$272937 \citep[optically classified
      L5;][]{Gizis2002} and 2MASS~J03185403$-$3421292 \citep[optically
      classified L7;][]{Kirkpatrick2008} were the best 
      matches for {\namesh}B. SpeX data for the templates are from
      \citet{Burgasser2010a} and A.~J.~Burgasser et al. (in preparation). 
      A $\chi^2$ weighted mean of all the best fit templates gives
      spectral types of M9.0$\pm$0.5 and L6$\pm$1 for the two
      components, respectively. The residuals of the comparison
      (target $-$ spectral type templates) are shown at the bottom of each panel.}
    \label{Fig: nirspec_temp}
  \end{centering}
\end{figure}

To determine spectral types we compared our NIR spectra of {\namesh}AB
with 463 spectra of 439 M7 or later dwarfs from the SpeX Prism
Spectral Libraries\footnote{\url{http://browndwarfs.org/spexprism/}}.
All templates wre chosen to have median S/N $>$ 10 and could not be
binaries, giants, subdwarfs, or spectral classifications that were
peculiar or uncertain. 
Best matches were determined by finding the minimum {\chisq} deviation
between component spectra and templates in the 0.95--1.35, 1.45--1.80, and
2.00--2.35~$\micron$ regions (i.e., avoiding telluric bands), following
the procedure of \citet{Cushing2008} with no pixel weighting.  The
two best matching templates to both components of {\namesh}AB are
shown in Figure~\ref{Fig: nirspec_temp}.  For {\namesh}A, these are
the optically classified M8 dwarf 2MASS~J05173729$-$3348593
\citep{Cruz2003} and L0 dwarf DENIS--P~J0652197$-$253450
\citep{Phan-Bao2008}; for {\namesh}B these are the optically
classified L5 dwarf~2MASSW~J1326201$-$272937 \citep{Gizis2002} and L7
dwarf 2MASS~J03185403$-$3421292 \citep{Kirkpatrick2008}. Note that
despite the differences in optical type, these spectra provide
equivalently good fits---the two fits were different only by
1.7$\sigma$ for the primary and 1.1$\sigma$ for the secondary based on
the F-test which gauges whether two different fits to data are
significantly distinct based on the ratio of {\chisq} values and
degrees of freedom \citep{Burgasser2010a}---to 
the components of {\namesh}AB, a reflection of the discrepancies
between optical and NIR spectral morphologies for late-type M and L
dwarfs \citep[e.g.,][]{Geballe2002, Kirkpatrick2005}.  A $\chi^2$
weighted mean of all the SpeX templates
\citep[e.g.,][]{Burgasser2010a} indicates component types of
M9.0$\pm$0.5 for {\namesh}A and L6$\pm$1 for {\namesh}B. 

We also derived classifications using a suite of spectral
indices and spectral index/spectral type relations from
\citet{Tokunaga1999}, \citet{Reid2001a}, \citet{Geballe2002},
\citet{Burgasser2006b}, and \citet{Burgasser2007c}. Table~\ref{Tab:
  indices} shows the measured values and the inferred spectral
subtypes of {\namesh}A and {\namesh}B for each of the indices. The
mean and scatter from these indices yields classifications of
L0.5$\pm$1.0 for {\namesh}A and L7.0$\pm$1.5 for {\namesh}B. These are
consistent with, but less precise than, the types inferred from
spectral template matching, so we adopt the latter for our subsequent
analysis.

\begin{deluxetable}{lcccccl}
  \tablecaption{Near-Infrared Spectral Indices}
  \tablewidth{0pt}
  \tablehead{
    \colhead{Index} & \multicolumn{2}{c}{2MASS}  & \colhead{} & \multicolumn{2}{c}{2MASS} & \colhead{Reference}\\ 
    \colhead{} & \multicolumn{2}{c}{J0130$-$4445A} & \colhead{} & \multicolumn{2}{c}{J0130$-$4445B} & \colhead{}\\ 
    \cline{2-3} 
    \cline{5-6}
    \colhead{} & \colhead{Value} & \colhead{SpT} & \colhead{} & \colhead{Value} & \colhead{SpT} & \colhead{}}
  \startdata
  {\wat}-J              & 0.954  &  L0.3   && 0.706 & L7.1    & 1, 2 \\ 
  {\wat}-H              & 0.876  &  M9.6   && 0.649 & L8.3    & 1, 2 \\ 
  {\wat}-A              & 0.686  &  L1.4   && 0.602 & L4.1    & 3    \\     
  {\wat}-B              & 0.818  &  L0.3   && 0.517 & L7.8    & 3    \\     
  {\wat}-1.5~$\micron$\tablenotemark{a}  & 1.216  &  M9.8   && 1.882 & L9.0    & 4    \\
  {\meth}-K             & 1.056  &  L2.1   && 0.968 & L5.5    & 1, 2 \\
  {\meth}-2.2~$\micron$\tablenotemark{b} & \nodata& \nodata && 1.033 & L5.7    & 4    \\
  K1\tablenotemark{c}   & 0.069  &   M8.7  && \nodata & \nodata    & 3, 5 \\
  \cline{1-7}
  Average SpT             &        &L0.5$\pm$1.0 &&       &L7.0$\pm$1.5 &      \\
  \enddata
  \label{Tab: indices}
  \tablerefs{
    (1) \citet{Burgasser2006b};
    (2) \citet{Burgasser2007c};
    (3) \citet{Reid2001a};
    (4) \citet{Geballe2002};
    (5) \citet{Tokunaga1999}}
  \tablenotetext{a}{The index {\wat}-1.5~$\micron$ is well-defined only
    for spectral types L0 or later.}
  \tablenotetext{b}{The index {\meth}-2.2~$\micron$ is well-defined only
    for spectral types L3 or later.}
  \tablenotetext{c}{The index K1 is well-defined only up to spectral
    type earlier than L6.}
\end{deluxetable}

\section{System Properties}\label{Sec: analysis}
\subsection{Is {\namesh} A Physical Binary?}
To assess whether two stars comprise a physical binary or are just a chance
alignment of random stars, the most reliable method used is to check for
a common systemic velocity. However, {\namesh}B is very faint, even in
the infrared, and has not been detected in any earlier epoch; hence, we
do not have proper motions for the secondary nor radial velocities for either
component. In the absence of kinematic information, we employed two
other tests to examine whether {\namesh}AB is a physical pair: 
(1) the heliocentric distances of the two components and (2) the
probability that the sources are a chance alignment based on the
surface distribution of stars on the sky.

The spectrophotometric distances to each component of {\namesh}AB were
derived using the $M_J$/spectral type relations from \citet{Cruz2003}
based on the combined-light 2MASS photometry and our relative SpeX
photometry (Table \ref{Tab: properties}). The derived distances are
34.5$\pm$3.2~pc for the primary and 45.8$\pm$13.6~pc for the
secondary, where the errors are from the uncertainties in the NIR
spectral types (see Sec. \ref{Sec: NIRanalysis}). These distances
are consistent with each other within their associated errors.

\begin{deluxetable}{lccl}
  \tablewidth{0pt}
  \tablecaption{Properties of {\name}AB}
  \tablehead{
    \colhead{Parameter} & \colhead{2MASS} & \colhead{2MASS} & \colhead{Reference}\\
    \colhead{} & \colhead{J0130-4445A} & \colhead{J0130-4445B} &\colhead{}}
  \startdata
  Optical Spectral Type              & M9           & \nodata  & 1 \\
  NIR Spectral Type & M9.0$\pm$0.5 & L6$\pm$1 & 2 \\
  $J$ (mag)\tablenotemark{a} & 14.12$\pm$0.03 & 17.28$\pm$0.06 & 2,3 \\ 
  $H$ (mag)\tablenotemark{a} & 13.48$\pm$0.03 & 16.13$\pm$0.10 & 2,3 \\ 
  $K_s$ (mag)\tablenotemark{a} & 12.99$\pm$0.03 & 15.34$\pm$0.05 & 2,3 \\ 
  {\JK} \tablenotemark{a}&  1.13$\pm$0.04 &  1.94$\pm$0.08 & 2,3 \\
  Est.\ {\Teff} (K)\tablenotemark{b} & 2400$\pm$110 & 1450$\pm$100 & 4 \\
  Est.\ Distance (pc) \tablenotemark{c}  & 35$\pm$3 & 46$\pm$14 & 2,5 \\ 
  \cline{1-4}
  Projected Separation (AU) \tablenotemark{d} & \multicolumn{2}{c}{130$\pm$50} & 1,2 \\
  {\vtan} ({\kms})\tablenotemark{d} & \multicolumn{2}{c}{23$\pm$6} & 2,6 \\
  \enddata
  \label{Tab: properties}
  \tablenotetext{a}{Calculated using our relative magnitudes and the
    combined-light 2MASS photometry for the system.}
  \tablenotetext{b}{Reported uncertainties include scatter in the
    {\Teff}/spectral type relation of \citet{Looper2008b} relation and
    uncertainties in the component classifications ($\pm$0.5 subtypes).} 
  \tablenotetext{c}{Based on the $M_J$/spectral type relation of
    \citet{Cruz2003}; uncertainties include photometric uncertainties
    and scatter in the \citeauthor{Cruz2003} relation.} 
  \tablenotetext{d}{Based on the average distance of 40$\pm$14~pc.}
  \tablerefs{
    (1) \citet{Reid2008}; 
    (2) This paper; 
    (3) 2MASS \citep{Cutri2003}; 
    (4) \citet{Stephens2009}; 
    (5) \citet{Cruz2003};
    (6) \citet{Faherty2009}.}
\end{deluxetable}

Next, we calculated the probability that {\namesh}AB is a random
chance alignment along our line-of-sight based on its
three-dimensional position in the Galaxy. We followed
\citet{Dhital2010}, constructing a three-component Galactic model with
the thin disk, thick disk, and halo, constrained with
empirical stellar density profiles \citep{Juric2008,
  Bochanski2010}. The model recreates a 30$\arcmin\times$30$\arcmin$
region in the sky, centered around the coordinates of the 
given binary system, and out to heliocentric distances of 2500~pc. As
all the simulated stars are single and non-associated, any visual
binary is a random chance alignment. In $10^7$ Monte Carlo
realizations, we found, on average, 0.0285 chance alignments per
realization on the sky, within the 3$\farcs$3 angular separation of
{\namesh}AB. More importantly, none of these chance alignments were
within the range of spectrophotometric distances (40$\pm$14~pc)
estimated for {\namesh}AB. As such, we conservatively infer a
$\lesssim 10^{-7}$ probability of positional coincidence. 
We therefore conclude that {\namesh}AB is a physically-bound binary
and not a chance alignment of two unassociated stars. 

\subsection{Age \& Mass Estimates for {\namesh}AB}
\begin{figure}
  \begin{centering}
    \includegraphics[width=1\linewidth]{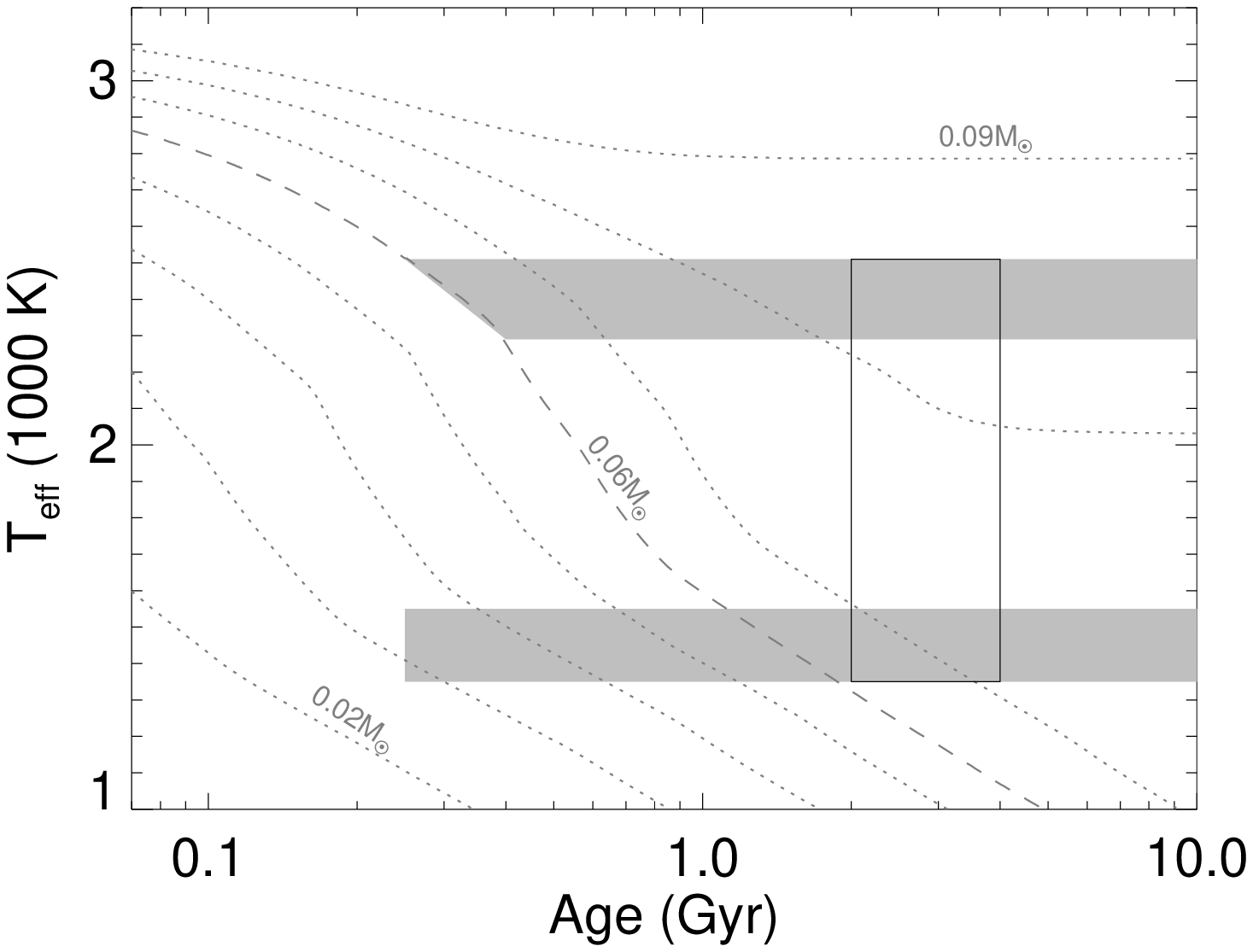}
    \caption{{\Teff} versus age for VLM stars and BDs based on
      the evolutionary models of \citet{Burrows1997}.  Tracks for
      masses of 0.02--0.09~{\Msun} in 0.01~{\Msun} steps are shown as
      dotted lines while the LDB limit of 0.06~{\Msun} is shown as a
      dashed line.  The locus defined by this lower limit on age and
      {\Teff} derived from their spectral types is shown in gray for
      both components. The solid rectangle shows the age estimate of
      2--4~Gyr based on the kinematics of \citet{Faherty2009}.
      Note that the broad range of possible ages for this system allow
      a wide range of mass ratios, from 0.6 to 0.9.}
    \label{Fig: evol_track}
  \end{centering}
\end{figure}

The NIR spectral types of {\namesh}A and {\namesh}B correspond to
effective temperatures, {\Teff}, of 2400$\pm$110~K and 1450$\pm$100~K,
respectively, based on the {\Teff}/spectral type relation of
\citet{Stephens2009}. Uncertainties include scatter in the {\Teff}
relation and uncertainties in the classifications (Sec. \ref{Sec:
  NIRanalysis}) added in quadrature. Figure~\ref{Fig: evol_track}
shows the \citet{Burrows1993, Burrows1997} evolutionary models,
displaying {\Teff} as a function of mass and age; the Burrows mass
tracks and the observed {\Teff} ranges are shown as dotted and dashed
lines and gray boxes, respectively. As the masses of the two
components vary quite a bit with assumed age, it is imperative to
constrain the age of {\namesh}AB. 

The absence of {\Lii} in {\namesh}A indicates that it is more massive
than the predicted lithium depletion boundary (LDB) mass,
$\sim$0.06~{\Msun} \citep{Chabrier1996, Burrows2001} for field
stars of solar metallicity. Using the {\Teff} for the primary based on
its spectral type (including uncertainties) and assuming a mass
$\gtrsim$0.06~{\Msun}, the evolutionary models of \citet{Burrows1997}
indicate an age $\gtrsim$250~Myr (Figure~\ref{Fig:
  evol_track}). We note that recent work by \citet{Baraffe2010} has
suggested that episodic accretion during the pre-main sequence stages
causes central temperature of a star to increase up to 1~dex, with a sharp
dependence on the frequency and magnitude of the episodic
accretion. This serves to deplete {\Lii} earlier than in
non-accreting stars of the same mass and effectively reduces the inferred LDB
mass and, thus, the  minimum allowable age. Here, we have not taken 
episodic accretion into account.
While we do not have an optical spectrum of the secondary, the
presence (absence) of {\Lii} in the spectrum would set a upper (lower)
limit on the mass and, hence, the age of the system, in this case
$\lesssim$1.8~Gyr ($\gtrsim$1.1~Gyr). The likely proximity of the mass
of the secondary to the LDB is motivation to obtain an optical
spectrum of this component.

The absence of {\ha} emission in the optical spectrum of
{\namesh}A and lack of UV or X-ray flux---the system is not detected
in the GALEX \citep{Martin2005b} or the ROSAT \citep{Voges1999} All-Sky
Surveys---indicates that the system is not particularly active and,
hence, not likely to be a very young system.  This absence indicates
that {\namesh}AB is probably older than 1--100~Myr, as such emission
has been detected in brown dwarfs in the Orion Nebula Cluster
\citep[isochronal age $\sim$1~Myr;][]{Peterson2008}, Taurus
\citep[$\sim$3~Myr;][]{Guieu2006}, $\sigma$ Orionis
\citep[2--7~Myr;][]{Zapatero-Osorio2002}, $\alpha$ Persei 
\citep[$\sim$80~Myr;][]{Stauffer1999}, Pleiades 
\citep[$\sim$100~Myr;][]{Stauffer1998, Martin2000}, and Blanco 1
($\sim$100~Myr; P.~A.~Cargile et al., in prep.). However, activity
signatures might not be reliable age indicators in the VLM
regime.  Both {\ha} and 
X-ray emission drop precipitously across the M dwarf/L dwarf
transition \citep[e.g.,][]{Kirkpatrick2000, Gizis2000, West2004,
  Stelzer2006}, likely the result of reduced magnetic field coupling
with increasingly neutral photospheres \citep[e.g.,][]{Gelino2002,
  Mohanty2002}.

Extreme youth can also be ruled out based on the the NIR
spectra of these sources, which do not exhibit the triangular H-band
peaks seen in $\sim$100~Myr Pleiades M and L dwarfs \citep{Bihain2010}
and young field L dwarfs \citep[e.g.,][]{Kirkpatrick2006}. It is
notable that {\namesh}B is somewhat red compared to typical L6 dwarfs
\citep[$\langle{J-K_s}\rangle =$ 1.82$\pm$0.07;][]{Schmidt2010}, as
red sources have been shown to exhibit smaller velocity 
dispersions and, hence, younger ages \citep{Faherty2009, Schmidt2010}.
However, {\namesh}A is not unusually red for its spectral type
($\langle{J-K_s}\rangle =$ 1.12$\pm$0.10); and the red 
color of the secondary may reflect an unusually dusty atmosphere
\citep[e.g.,][]{Looper2008a}.  Also, neither NIR spectra nor the
optical spectrum of {\namesh}A show high gravity signatures, i.e.,
unusually blue colors form enhanced {\Hh},  or evidence of the system
being metal-poor, making it unlikely that the system is as old as
$\sim$10~Gyr \citep{Burgasser2003b, Reid2007}. 

Considering the kinematics of the system, the tangential velocity of
{\namesh}A, 19$\pm$3~{\kms} is similar to
the median velocities of the L dwarfs in the SDSS sample  
\citep[28$\pm$25~{\kms};][]{Schmidt2010}, the M9 dwarfs in the BDKP
sample \citep[23$\pm$23~{\kms};][]{Faherty2009}, and the M7--L8
dwarfs in the 2MASS sample \citep[25$\pm$21~{\kms};][]{Schmidt2007},
with the quoted errors being the 1$\sigma$ dispersions. The low tangential
velocity suggests that {\namesh}AB is part of the thin disk, although we
note that we cannot rule out a higher space velocity for the binary system.
Kinematic studies have found that late-M and L dwarfs with average
kinematics are typically $\sim$2--4~Gyr old \citep{Wielen1977, Faherty2009}.

In conclusion, based on the absence of {\Lii} in the primary, we can
place a (model-dependent) hard limit on the minimum age of {\namesh}AB
to be $\sim$250~Myr while its kinematics indicate a preferred age of
$\sim$2--4~Gyr. Spectral features for both components are in 
agreement with these ages. For the ages of 0.25--10~Gyr, based on the
\citet{Burrows1993} and \citet{Burrows1997} models, the estimated
masses for {\namesh}A and {\namesh}B are 0.055--0.083 and
0.032--0.076~{\Msun}, respectively, and the mass ratio is
0.57--0.92. For kinematics-based age limits of 2--4~Gyr, the estimated
masses for {\namesh}A and {\namesh}B are 0.082--0.083 and
0.066--0.073~{\Msun}, respectively, and the mass ratio is 0.81--0.89
(Table ~\ref{Tab: model_props}). Hence, the components straddle the
hydrogen-burning mass limit; and this system is likely composed of a
very low mass star and (massive) brown dwarf pair.

\begin{deluxetable}{lllll}
  \tablewidth{0pt}
  \tablecaption{Model-dependent Properties of {\namesh}AB \citep{Burrows1993, Burrows1997}}
  \tablehead{
    \colhead{Parameter} & \multicolumn{4}{c}{Age (Gyr)} \\
    \cline{2-5}
    \colhead{} & \colhead{0.25} & \colhead{2} & \colhead{4} &\colhead{10}}
  \startdata
  Primary Mass ({\Msun})   & 0.055 & 0.082 & 0.083 & 0.083 \\
  Secondary Mass ({\Msun}) & 0.032 & 0.066 & 0.073 & 0.076 \\
  Mass Ratio               & 0.57  & 0.81  & 0.89  & 0.92  \\
  Log Binding Energy (erg) & 41.61 & 41.96 & 41.97 & 41.97 \\
  Period (yr)              & 5030  & 3860  & 3760  & 3720  \\
  \enddata
  \label{Tab: model_props}
\end{deluxetable}

\section{Discussion}\label{Sec: discussion}
\subsection{Formation of Wide VLM Binaries in the Field}
With a projected separation of 130$\pm$50~AU, {\namesh}AB is one of
only ten VLM systems wider than 100~AU, with six of them in the
field. All of these systems have been identified relatively
recently; prior to their discovery, it was believed that VLM field
systems were nearly all tight, a possible consequence of dynamic
ejection early on. Based on this idea and the VLM binary population
known at the time, two relations to define the largest possible
separation of VLM binaries were proposed. First,  
\citet{Burgasser2003a} suggested that the maximum separation of a
system was dependent on its mass: $a_{\rm max}$ (AU) $=$ 1400
({\Mtot}/{\Msun})$^2$. Second, \citet{Close2003} proposed 
that the stability of binary systems was contingent on their binding
energy---a criterion based on the product rather than the sum of
component masses; thus, only systems with binding energy $\geq
10^{42.5}$~erg would exist in the field\footnote{We note that for the
  small separations and a mass ratio highly skewed toward one, which
  was the case for the VLM binaries known at the time, the
  \citet{Burgasser2003a} and \citet{Close2003} limits are essentially
  equivalent.}. For the (age-dependent) estimated mass of {\namesh}AB, the
\citet{Burgasser2003a} relation equates to maximum physical separations
of only 9.2~AU and 35.4~AU for ages of 0.25 and 10~Gyr, respectively,
which are both much smaller than the physical separation we have
measured for {\namesh}AB. Similarly, the binding energies for the
system are $10^{41.61}$ and $10^{41.97}$~erg for the same ages (see
Table~\ref{Tab: model_props}). Both the \citet{Burgasser2003a} and 
\citet{Close2003} relations are definitively violated by {\namesh}AB,
for all ages and mass ratios. Assuming these limits emerge from
dynamical scattering processes, this binary seems unlikely to have
formed via the ejection of protostellar embryos.

More recently, \citet{Zuckerman2009} have argued that
fragmentation, rather than dynamical, processes are more likely to
describe the boundary for the lowest binding energy systems. A
protostellar cloud can only fragment if it is more massive than the
minimum Jeans mass \citep[$\sim$7~M$_{\rm J}$;][]{Low1976}. Assuming
fiducial separations of 300~AU for the fragments, they derived a
cut-off for binding energy as a function of total systemic
mass. Finding that this disfavors the formation of very wide and/or
high mass ratio binaries, \citet{Faherty2010} used the Jeans length,
instead of the fiducial separation, and mass ratio of the system.
For {\namesh}AB at 0.25~Gyr ({\Mtot}$\approx$0.1~{\Msun}),
\citet{Zuckerman2009} and \citet{Faherty2010} relations suggest
minimum binding energies of 10$^{40.5}$ and 10$^{39}$~erg,
respectively; if the system were older, they would be even more stable
due to the higher masses. {\namesh}AB is well within the bounds of both 
\citet{Zuckerman2009} and \citet{Faherty2010} formation criteria for
all ages and mass ratios. Hence, the observed wide, low binding energy
VLM binaries could have formed from small protostellar clouds, with
masess close to the local Jeans mass.

Current numerical simulations have suggested an alternative mechanism
to form wide binaries: N-body dynamics in small clusters disrupt
the VLM pairs wider than $\sim$60~AU, and very wide systems
($>10^4-10^5$~AU) can then be formed when stars are ejected into the
field in the same direction \citep{Moeckel2010, Kouwenhoven2010}. 
While this provides a mechanism to create the most fragile VLM pairs
identified to date, it does not aid in the formation of 100--1000~AU pairs
like {\namesh}AB. Two other VLM systems in this separation range are known:
2MASSJ~1623361$-$240221 \citep[220~AU;][]{Billeres2005} and
SDSS~J141623.94$+$134826.3 \citep[100~AU;][]{Burningham2010, Scholz2010}.

\subsection{{\namesh}AB as a Probe of the M Dwarf/L Dwarf Transition}\label{Sec: context}
The components of {\namesh}AB straddle a spectral type range
that is particularly interesting for three reasons.  First, condensate
clouds become an important source of photospheric opacity and thermal
evolution starting at the end of the M dwarf sequence and peaking in
influence in the middle of the L dwarf sequence \citep{Ackerman2001,
  Kirkpatrick2008, Saumon2008}. With a component on either end of this
regime, {\namesh}AB is a particularly useful coeval laboratory for
studying the emergence and dispersal of these clouds.  Second, both
components straddle the hydrogen-burning mass limit; and the secondary
is very near the LDB.  Detection of {\Lii} in the optical
spectrum of the secondary could provide a relatively precise 
constraint on the age of this system (0.25--1.8~Gyr) and thereby make
it a useful benchmark for studies of brown dwarf thermal evolution and
atmospheric models \citep[e.g.,][]{Pinfield2006}. Third, the M
dwarf/L dwarf transition exhibits a steep decline in magnetic activity
metrics, including {\ha}, UV, and X-ray emission
\citep[e.g.,][]{Gizis2000, West2004} but notably not radio emission
\citep[e.g.,][]{Berger2006, Berger2010}.  This is believed to be due to the
decoupling of magnetic fields from an increasingly neutral photosphere
\citep{Gelino2002, Mohanty2002} but does not rule out the
presence of significant magnetic fields \citep{Reiners2007}.
While {\ha} is absent in the spectrum of {\namesh}A,
examination of field strengths and radio emission in this coeval pair
may facilitate understanding of how magnetic fields evolve across the
stellar/brown dwarf transition.  As one of only three binaries
spanning the M dwarf/L dwarf transition whose components are easily
resolvable from ground-based facilities \citep[the other two are the
1$\farcs$2 L1.5$+$L4.5 2MASSJ~1520022$-$442242 and the 1$\farcs$0
M9$+$L3 2MASSJ~1707234$-$055824;][]{Burgasser2004a, Burgasser2007a, 
  Folkes2007}, {\namesh}AB is an important laboratory for 
studying how condensate clouds, lithium burning, and magnetic activity
trends vary across this transition. 

\section{Summary}\label{Sec: summary}
We have identified a binary companion to {\name}A based on NIR imaging
and spectroscopic observations. The secondary is well-separated
($\Delta\theta=3\farcs2$) and much fainter ($\Delta K \approx$
2.35~mags). Based on template matching and spectral
indices, we have calculated the NIR spectral types to be M9.0$\pm$0.5
and L6$\pm$1 for the two components. The optical spectrum of
{\namesh}A shows no evidence of either {\ha} or {\Lii} indicating a
minimum age of 0.25~Gyr, while the kinematics suggest an age of
2--4~Gyr. However, we would like to stress that 0.25--10~Gyr, with the
lower bound set by LDB and activity, is the more secure age for the
system. {\namesh}AB is likely a ``grown-up'' wide binary that has
survived ejections and/or dynamical interactions. More importantly,
the system definitively violates the binary stability limits based on
the ejection hypothesis \citep{Burgasser2003a, Close2003} and satisfies
the limits based on the idea that wide VLM binaries are formed
from approximately Jeans mass-sized protostellar clouds
\citep{Zuckerman2009, Faherty2010}. This suggests that observed wide
VLM binaries may have formed differently than single VLMs and/or
tighter binaries. As one of ten VLM systems with separations
$\gtrsim$100~AU, {\namesh}AB provides a stringent test for theoretical
studies of VLM binary formation, as well as a well-resolved, coeval
laboratory for studying empirical trends across the M dwarf/L dwarf
and stellar/brown dwarf transitions.

\acknowledgments
The authors would like to thank the anonymous referee for
his/her comments on the manuscript, telescope operator Paul Sears 
and instrument specialist John Rayner for their assistance during the
IRTF observations, and Kelle Cruz for providing an electronic version of
the optical spectrum of {\namesh}A. SD and KGS acknowledge funding
support through NSF grant AST-0909463. This publication makes use of data 
from the Two Micron All Sky Survey, which is a joint project of the
University of Massachusetts and the Infrared Processing and Analysis
Center, and funded by the National Aeronautics and Space
Administration and the National Science Foundation. 2MASS data were
obtained from the NASA/IPAC Infrared Science Archive, which is
operated by the Jet Propulsion Laboratory, California Institute of
Technology, under contract with the National Aeronautics and Space
Administration. This research has made use of the VLM Binaries
Archive, maintained by Nick Siegler at \url{http://www.vlmbinaries.org};  
the SpeX Prism Spectral Libraries, maintained by Adam Burgasser at
\url{http://www.browndwarfs.org/spexprism}; and the M, L, and T dwarf
compendium housed at DwarfArchives.org and maintained by Chris Gelino,
Davy Kirkpatrick, and Adam Burgasser. The authors wish to recognize
and acknowledge the very significant cultural role and reverence that
the summit of Mauna Kea has always had within the indigenous Hawaiian
community.  We are most fortunate to have the opportunity to conduct
observations from this mountain. 

Facilities: \facility{IRTF (SpeX)}

\bibliography{ads}
\end{document}